\documentclass[
    amsmath,
    amssymb,
    english,
    showkeys,
]{revtex4-2}
\usepackage{fix-cm}
\usepackage[utf8]{inputenc}
\usepackage[english]{babel}%portuguese,
\usepackage{graphicx}
\usepackage{latexsym,amsthm,amsfonts}
\usepackage{makeidx}   
\usepackage[T1]{fontenc}
\usepackage{subfigure}
\usepackage{dcolumn}
\usepackage{bm}
\usepackage{cancel}
\usepackage{xcolor}
\usepackage[hyperindex,colorlinks,linkcolor=blue,citecolor=red]{hyperref}

\usepackage{textgreek}
\usepackage{comment}

\allowdisplaybreaks[1]

\usepackage{upgreek}
\usepackage{graphicx}
\usepackage{tensor}

\begin{document}
\title{Black-bounce solution in $k$-essence theories}

\author{Carlos F. S. Pereira}
	\email{carlos.f.pereira@edu.ufes.br}
	\affiliation{Departamento de Física, Universidade Federal do Espírito Santo, Av. Fernando Ferrari, 514, Goiabeiras, 29060-900, Vit\'oria, ES, Brazil.}
	
	\author{Denis C. Rodrigues}
    \email[]{deniscr@gmail.com}
	\affiliation{Núcleo Cosmo-ufes \& Departamento de Física, Universidade Federal do Espírito Santo, Av. Fernando Ferrari, 514, Goiabeiras, 29060-900, Vit\'oria, ES, Brazil.}
		
	\author{Júlio C. Fabris}
    \email[]{julio.fabris@cosmo-ufes.org}
	\affiliation{Núcleo Cosmo-ufes \& Departamento de Física, Universidade Federal do Espírito Santo, Av. Fernando Ferrari, 514, Goiabeiras, 29060-900, Vit\'oria, ES, Brazil}
    \affiliation{National Research Nuclear University MEPhI (Moscow Engineering Physics Institute), 115409, Kashirskoe shosse 31, Moscow, Russia.}
	
	\author{Manuel E. Rodrigues}
    \email[]{esialg@gmail.com}
	\affiliation{Faculdade de Ci\^encias Exatas e Tecnologia, Universidade Federal do Par\'a Campus Universit\'ario de Abaetetuba, 68440-000, Abaetetuba, Par\'a, Brazil and Faculdade de F\'isica, Programa de P\'os-Gradua\c{c}\~ao em F\'isica, Universidade Federal do Par\'a, 66075-110, Bel\'em, Par\'a, Brazil.}
\begin{abstract}

\noindent In the present work, we construct black-bounce configurations in the context of $k$-essence theory. The solutions have a regular metric function at the origin. The area metric function is linked to the black-bounce area initially considered by Simpson–Visser, $\Sigma^2=x^2+a^2$. Subsequently, the expressions for the scalar field and scalar potential corresponding to the found solutions are determined, exhibiting phantom behavior everywhere due to violation of Null Energy Condition $(NEC^\phi)$. The Kretschmann scalar is regular throughout spacetime, and the geodesics are complete. The energy conditions are analyzed, verifying that the null $(NEC^\phi_1)$ and dominant energy conditions $(DEC^\phi_1)$ are violated inside and outside the event horizon. Finally, the extrinsic curvature method was applied to determine the sign of the mass on the junction surface.
\end{abstract}
	
\keywords{Black-bounce, $k$-essence theory, energy conditions}
	
\maketitle
	
\section{Introduction}\label{sec1}

Recently, Simpson and Visser \cite{matt} introduced a new class of solutions called “black-bounce” describing regular black holes and traversable wormholes. These solutions have a non-zero throat radius $a^2 \neq 0$ and reduce to the Schwarzschild metric when $a \to 0$. Subsequent works have explored generalizations and applications of the black-bounce solutions. Lobo et al. \cite{lobo} constructed new black-bounce solutions by modifying the mass function, recovering the original Simpson-Visser solution \cite{matt} for particular parameter values. Rodrigues and Silva \cite{manoel1} investigated the Simpson-Visser black-bounce metric with modifications to the metric function related to the black-bounce area. Junior and Rodrigues \cite{manoel2} obtained novel black-bounce solutions in the context of $f(T)$ modified gravity theory.

The search for exotic solutions like regular black holes and traversable wormholes requires violating standard energy conditions: minimally coupled canonical scalar field cannot describe such geometries. However, Bronnikov and Fabris showed a canonical scalar field with phantom behavior can allow regular black holes \cite{phantom}. In this context, $k$-essence theory has emerged as an exotic matter alternative, with its non-canonical kinetic term displaying phantom behavior without exotic matter. $k$-Essence theories generalize the scalar field kinetic term, originally proposed for modeling primordial inflation with just a kinetic term \cite{mukha1, mukha2,mukha3}. Generalized kinetic terms are also motivated by string theory \cite{dbi}. This work examines black-bounce solutions in $k$-essence theory with a power law kinetic term and potential, focusing on energy condition violations.

In the studies of static, spherically symmetric configurations, exotic matter is frequently introduced in order to find regular black holes and wormholes solutions in nonlinear electrodynamics. These new regular metrics constitute exact solutions in general relativity, derived through a combined stress-energy tensor of a scalar field with non-zero self-interaction potential and a magnetic field \cite{nlinear1,nlinear2,nlinear3,nlinear4,nlinear5}. However, rotating metrics have also been found to accommodate such regular objects \cite{rotation1,rotation2,rotation3}. This analysis investigates black-bounce solutions in $k$-essence theory to gain insights into $k$-essence and exotic solutions in general relativity.  

Futhermore, Bronnikov et al. \cite{simple} explored Ellis-Bronnikov wormhole solutions in extended gravity theories. The analysis shows the same wormhole metric emerges in Rastall gravity and k-essence theories but with different stability properties. Perturbation analysis reveals inconsistencies in Rastall gravity, while the $k$-essence solution is unstable for certain model parameters. The results highlight challenges in finding simple, traversable, and perturbatively stable wormhole solutions without exotic matter.

The Simpson-Visser metric has been studied in other contexts, such as light deflection and lensing effects \cite{adriano1}. Gravitational lensing was analyzed using black-bounce solutions in a spherically symmetric and stationary spacetime \cite{bouncerotation1, bouncerotation2}. In the zero mass limit, this reduces to the Ellis-Bronnikov charged wormhole. Quantum dynamics have been studied using the Simpson-Visser metric \cite{adriano2,adriano3,adriano4,outro}.

Phantom scalar fields are often studied as a source of exotic matter required to obtain wormhole solutions minimally coupled to general relativity \cite{nlinear5,N1}. Their phantom properties are typically associated with violating energy conditions and sometimes instabilities \cite{N2,N3}. Additionally, ghost fields are commonly associated with dark energy candidates, further emphasizing the importance of investigations in this direction \cite{N4,N5}. From this perspective, phantom fields have been explored as a matter source for singular \cite{N6,N7} and regular black holes \cite{phantom,N8,N9}.
%\textcolor{blue}{Phantom scalar fields are often studied as a source of exotic matter necessary to obtain wormhole solutions minimally coupled to general relativity \cite{nlinear5,N1}. Its phantom characteristic is generally associated with violation of energy conditions and in some cases with instabilities \cite{N2,N3}. In addition to all this, ghost fields are generally associated with dark energy candidates, further strengthening the importance of investigations in this sense \cite{N4,N5}. From this perspective, phantom fields have been investigated as a source of matter for singular \cite{N6,N7} and regular black holes \cite{phantom,N8,N9}.} 

The paper first establishes in Section \ref{sec2} the theoretical background of the $k$-essence model, including the key relationships and equations. Section \ref{sec3} then derives the specific metric function corresponding to a defined black-bounce throat geometry, and determines the associated scalar field and potential solutions that satisfy the equations of motion. Next, Section \ref{sec4}, examines the geometric properties by defining the regular Kretschmann scalar and stress-energy tensor components inside and outside the horizon, as well as analyzing the  energy conditions required for the black-bounce solutions. Finally, in Section \ref{sec5}, summarizes the main conclusions from this analysis regarding the viability of constructing regular black-bounce geometries within $k$-essence theories.

\section{General relations}\label{sec2}

$k$-Essence theories are characterized by a non-canonical kinetic term for the scalar field, represented by the Lagrangian
\begin{equation}\label{Lagran}
    \mathcal{L} = \sqrt{-g}[R-F(X,\phi)]\,,
\end{equation}
where $R$ is the Ricci scalar and $X=\eta\phi_{;\rho}\phi^{;\rho}$ denotes the kinetic term. While $k$-essence models can include a potential term and non-trivial couplings, the scalar sector is generally minimally coupled to gravity. The parameter $\eta=\pm 1$ avoids imaginary terms in the kinetic expression $X$. By choosing different forms of the function $F(X,\phi)$, $k$-essence theories can describe both phantom and standard scalar fields.

The variation of the Lagrangian \eqref{Lagran} with respect to the metric tensor and the scalar field yields the field equations.
\begin{eqnarray}\label{eq1}
G_\mu^{\nu}=-T_\mu^{\nu}\left(\phi\right)=-\eta{F_X}\phi_\mu{\phi^{\nu}} + \frac{1}{2}\delta_\mu^{\nu}F, \\\label{eq2}
\eta\nabla_\alpha\left(F_X\phi^{\alpha}\right)-\frac{1}{2}F_\phi =0,
\end{eqnarray} 
where $G_\mu^{\nu}$ is the Einstein tensor, $T_\mu^{\nu}$ the stress-energy tensor, $F_X=\frac{\partial{F}}{\partial{ X}}$, $F_\phi=\frac{\partial{F}}{\partial\phi}$ and $\phi_\mu=\partial_\mu\phi$.

The line element representing the most general spherically symmetric and static spacetime takes the form:
\begin{eqnarray}\label{eq3}
ds^2=e^{2\gamma\left(u\right)}dt^2-e^{2\alpha\left(u\right)}du^2-e^{2\beta\left(u\right)}d\Omega^2,
\end{eqnarray} where $u$ is an arbitrary radial coordinate, $d\Omega^2=d\theta^2+\sin^2\theta{d\varphi^2}$ the volume element, and $\phi=\phi\left (u\right)$. 

The non-zero components of the stress-energy tensor are,
\begin{eqnarray}\label{eq4}
T_0^{0}= T_2^{2} =T_3^{3}= -\frac{F}{2}, \\\label{eq5}
T_1^{1}=-\frac{F}{2} -\eta{F_X}e^{-2\alpha}{\phi}'^2,
\end{eqnarray} with $\phi'=\frac{d\phi}{du}$.

It is assumed that the function $X=-{\eta}e^{-2\alpha}{\phi}'^2$ is positive, which implies that $\eta=-1$. As a result, the equations of motion take the form:
\begin{eqnarray}\label{eq6}
2\left(F_X{e^{-\alpha+2\beta+\gamma}}\phi'\right)' - {e^{\alpha+2\beta+\gamma}}F_\phi=0, \\\label{eq7}
{\gamma}'' + {\gamma}'\left(2{\beta}'+ {\gamma}'-{\alpha}'\right)-\frac{e^{2\alpha}}{2}\left(F-XF_X\right)=0, \\\label{eq8}
-e^{2\alpha-2\beta} + {\beta}'' +{\beta}'\left(2{\beta}'+ {\gamma}'-{\alpha}'\right) -\frac{e^{2\alpha}}{2}\left(F-XF_X\right)=0, \\\label{eq9}
-e^{-2\beta} + e^{-2\alpha}{\beta}'\left({\beta}'+2{\gamma}'\right) -\frac{F}{2} + XF_X=0.
\end{eqnarray}

The notation used here follows the same as used in the reference \cite{bronnikov1}. The following coordinate transformation is defined: $u=:x$, and the \textit{guasiglobal} gauge $\alpha\left(u\right)+\gamma\left(u\right)=0$ is employed. As a result, the line element in Eq. (\ref{eq3}) can be expressed in the following form:
\begin{eqnarray}\label{eq10}
ds^2= A\left(x\right)dt^2- \frac{dx^2}{A\left(x\right)} - \Sigma^2\left(x\right)d\Omega^2,
\end{eqnarray}
where the metric functions are defined as $A(x) = e^{2\gamma} = e^{-2\alpha}$ and $e^\beta = \Sigma(x)$. The equations of motion defined in Eqs. (\ref{eq6}-\ref{eq9}) can then be rewritten in the new coordinates. Combining Eqs. (\ref{eq7}-\ref{eq9}) yields the expressions:
\begin{eqnarray}\label{eq11}
2A\frac{{\Sigma}''}{\Sigma} - XF_X =0, \\\label{eq12}
{A}''\Sigma^2 - A\left(\Sigma^2\right)''+ 2 =0,
\end{eqnarray} 
where the primes now represent derivatives with respect to $x$.

The two remaining equations, Eq. (\ref{eq6}) and Eq. (\ref{eq9}), are rewritten in the new coordinates as
\begin{eqnarray}\label{eq13}
2\left(F_X{A\Sigma^2}\phi'\right)' - \Sigma^2F_\phi = 0, \\\label{eq14}
\frac{1}{\Sigma^2}\left(-1 + A'\Sigma'\Sigma + A{\Sigma'}^2\right) -\frac{F}{2} + XF_X = 0.
\end{eqnarray}

\section{General solution} \label{sec3}

The analysis aims to find black-bounce solutions to the $k$-essence equations of motion \cite{livro,matt}. The metric function $\Sigma^2(x) = x^2 + a^2$ from the original work \cite{matt} is used, where the nonzero throat radius $a$ gives regular black holes or wormholes, with the area function $\Sigma^2(x)$ and the $k$-essence equations of motion, Eq. \eqref{eq12}, the corresponding metric function $A(x)$ is derived.

The general solution of the differential equation Eq. (\ref{eq12}) is given by
\begin{eqnarray}\label{eq14a}
A\left(x\right)= 1+ C_1\left[\left(x^2+a^2\right)\arctan\left(\frac{x}{a}\right) + xa\right] + C_2\left(x^2+a^2\right),
\end{eqnarray} 
where $C_1$ and $C_2$ are constants.

Certain requirements were imposed on the solution Eq. (\ref{eq14a}), such as being asymptotically flat, leading to a constraint between the constants $C_2 = -\frac{\pi}{2}C_1$. Furthermore, the solution should approach the Simpson-Visser solution as $x \to 0$, namely, $A(x \to 0) = 1 - \frac{2m}{a}$. Hence, the constant is set as $C_1 = \frac{4m}{\pi a^3}$. The resulting solution is:
\begin{eqnarray}\label{eq15}
A\left(x\right)=1 + \left(\frac{4m}{\pi{a^3}}\right)\left[xa + \left(x^2+a^2\right)\left(\arctan\left(\frac{x}{a}\right)-\frac{\pi}{2}\right)\right].
\end{eqnarray}

Figure \ref{fig1} shows curves of the metric function from Eq. (\ref{eq15}) for various throat radii $a$, inside and outside the event horizon. For all $a$, $A(x)$ diverges as $x \rightarrow -\infty$ and is asymptotically flat as $x \rightarrow \infty$. This general solution of Eq. (\ref{eq15}) is regular at the origin and for $x \rightarrow -\infty$, asymptotically approaching to de Sitter-Schwarzschild form. This requires considering the series expansion of $\arctan\left(\frac{x}{a}\right)$ for $x\rightarrow -\infty$ and discarding higher order terms $\mathcal{O}\left(\frac{1}{x}\right)$. Taking the general metric function in Eq. (\ref{eq15}) gives:
\begin{eqnarray}\label{eq15a}
A\left(x\right)= 1-\frac{8m}{3\pi}\left(\frac{1}{x}\right)-\frac{4m}{a^3}\left(x^2+a^2\right).
\end{eqnarray}

The general metric function in Eq. (\ref{eq15}) is equivalent to the solution in Eq. (10) from \cite{phantom}, with redefinitions $\rho_0=\frac{4m}{\pi}$ and $c=-\frac{2m}{a}$. This corresponds to the canonical $n=1$ phantom scalar field case in $k$-essence theory. The regularity of the general solution in Eq. (\ref{eq15}) can be seen in the Kretschmann scalar (\ref{KSDESITTER}), which tends to zero as $x\rightarrow\infty$ (Minkowski limit) and is constant and positive as $x\rightarrow-\infty$.

The behavior of the scalar field for the obtained $k$-essence solution, with $n=\frac{1}{3}$, can be examined using the general metric solution in Eq. (\ref{eq15}). The scalar field $\phi(x)$ for this metric is given by 
\begin{eqnarray}\label{eqphiDsitter}
\phi\left(x\right)= \frac{D_1}{4a^5}\left[\frac{xa^3}{\Sigma^4}+\frac{3xa}{2\Sigma^2}+\frac{3}{2}\arctan\left(\frac{x}{a}\right)\right] - \frac{D_1m}{\pi{a^2}\Sigma^4} - \frac{D_1m}{a^6}\left[\frac{ax}{\Sigma^2}+\arctan\left(\frac{x}{a}\right)\right] \\\nonumber
+ \left(\frac{2D_1m}{\pi{a^6}}\right)\left[\frac{a^2}{2\Sigma^2}+ \arctan\left(\frac{x}{a}\right)\left(\frac{xa}{\Sigma^2} + \frac{1}{2}\arctan\left(\frac{x}{a}\right)\right)\right],
\end{eqnarray} where $D_1=\left(\frac{6a^2}{F_0}\right)^\frac{3}{2}$ is a constant.

As shown in Figure \ref{ESCALARSDESITTER}, $\phi(x)$ approaches constant values depending on the throat radius $a$ as $x\rightarrow\pm\infty$, specifically
\begin{align}
\phi(x\rightarrow-\infty)=-\frac{9\pi\sqrt{\frac{3}{2}}}{4a^3}(a-4m)\quad \mbox{and}\quad
\phi(x\rightarrow\infty)=\frac{3\pi\sqrt{\frac{3}{2}}}{4a^3}(3a-4m)
\end{align}
where we set $F_0=1$.

Similarly, the potential $V(x)$ can be analyzed. The potential for the metric in Eq. (\ref{eq15}) is given by 
\begin{eqnarray}\label{eqpotN}
V\left(x\right)= \frac{2a^2}{\Sigma^4} - \frac{cax}{\Sigma^4}\left(\Sigma^2+2x^2\right) - \frac{c}{\Sigma^2}\left(3x^2-a^2\right)\left[\arctan\left(\frac{x}{a}\right)- \frac{\pi}{2}\right],
\end{eqnarray}  where $c=\frac{4m}{\pi{a^3}}$ is a combination of constants.

As Figure \ref{POTENCIALSDESITTER} exhibits, $V(x)$ tends to the constant $3\pi c$ as $x\rightarrow-\infty$ and to zero as $x\rightarrow\infty$.

\subsection{Black-Bounce solution}\label{sec31}

In order to construct black-bounce solutions, the general solution in Eq. (\ref{eq15}) will be matched to construct the appropriate geometry.

\begin{figure}[htb!]
		\centering
		\mbox{\subfigure[]
			{\label{fig1}
				\includegraphics[scale=0.4]{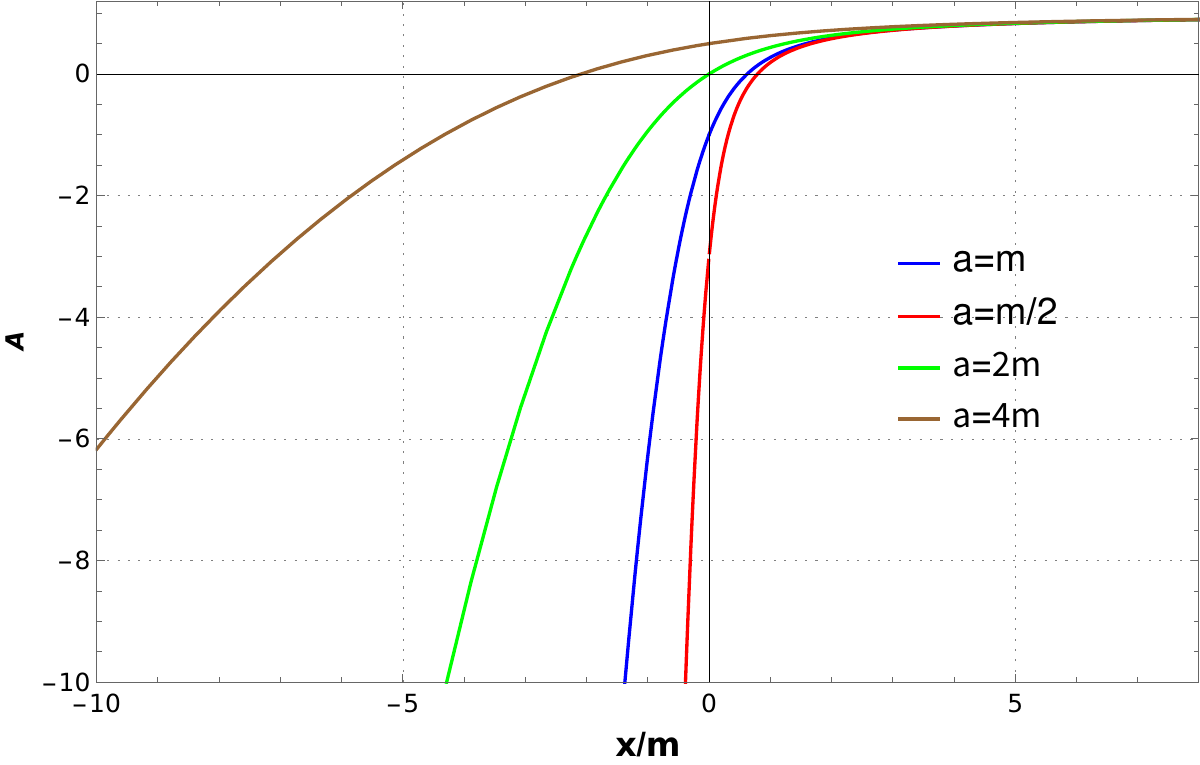}}\qquad
			\subfigure[]
			{\label{fig2}
				\includegraphics[scale=0.4]{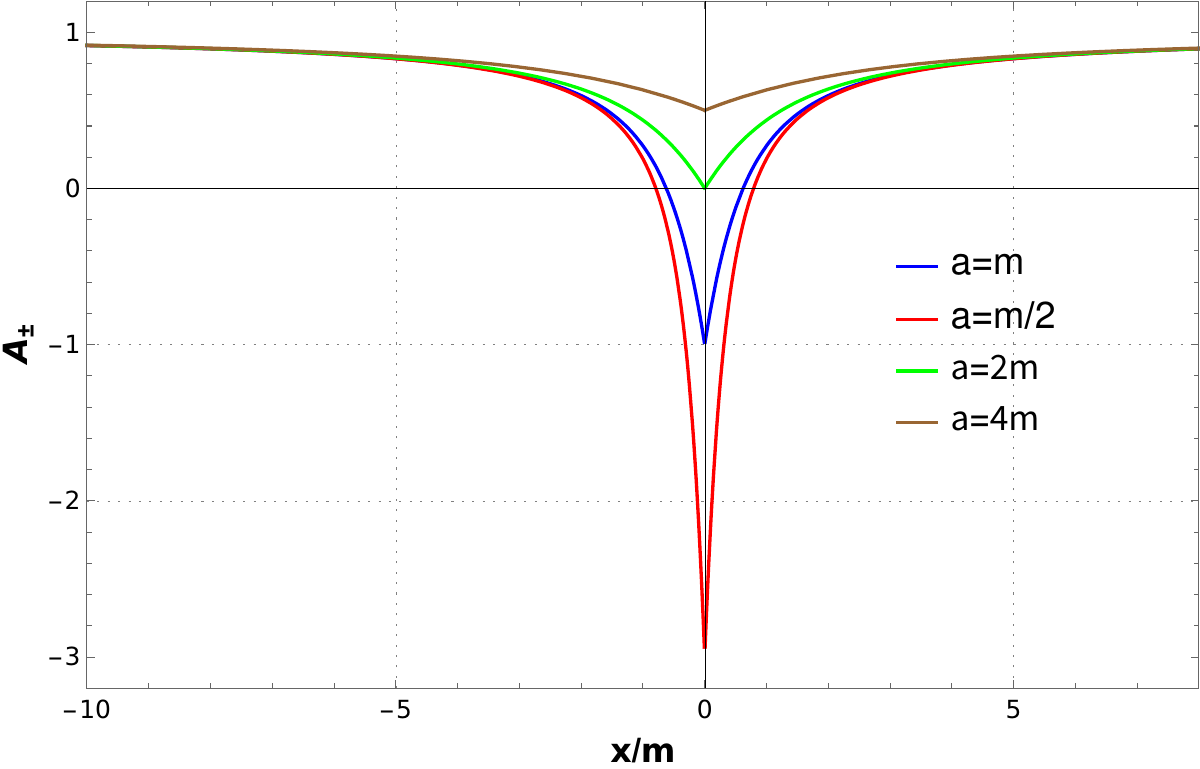}}
}
    \caption{(a) shows curves for various throat radius values $a$; the function is not asymptotically flat in both $x \to \pm \infty$ limits. (b) shows radii inside and outside the horizon, with the metric function defined by matching asymptotically flat solutions at $x = 0$ for $x \to \pm \infty$.}
\label{functions}
\end{figure}

\begin{figure}[htb!]
		\mbox{\subfigure[]
			{\label{POTENCIALSDESITTER}
				\includegraphics[scale=0.4]{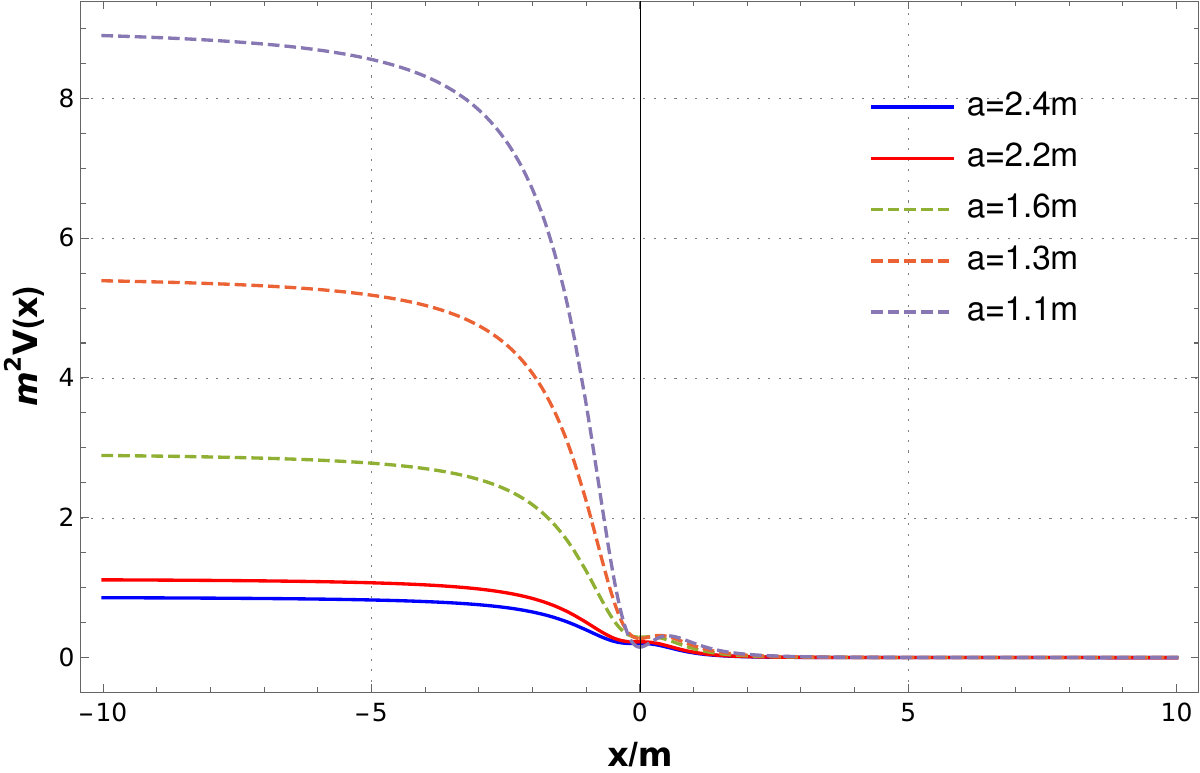}}\qquad
			\subfigure[]
			{\label{ESCALARSDESITTER}
				\includegraphics[scale=0.4]{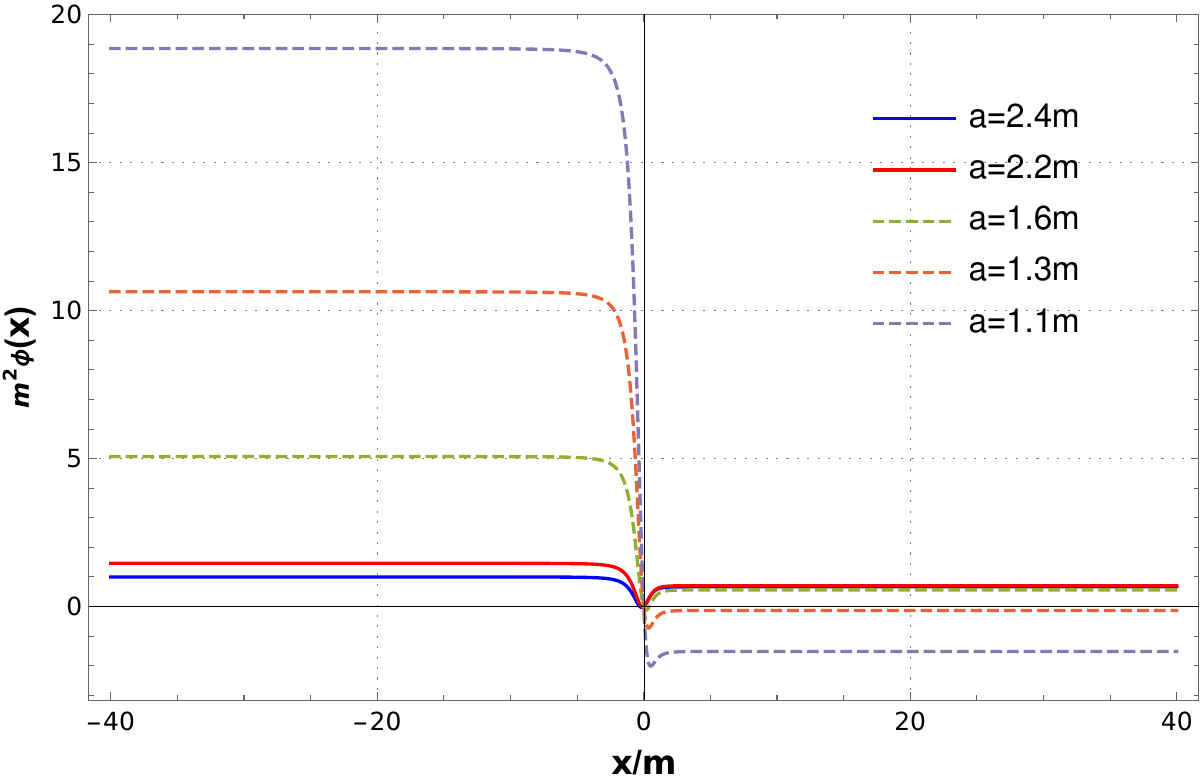}}
}
    \caption{Graphing for the scalar field and potential for the general metric function Eq. (\ref{eq15}) with radius values of throats outside the inside event horizon. We fixed the constant $F_0=1$.}
\label{POTESCALARDESITTER}
\end{figure}

First, the requirement was imposed that the metric function be asymptotically flat in both limits, to recover the Schwarzschild metric. To achieve this, the metric function Eq. (\ref{eq15}) was bisected at $x=0$ and mirrored, defining two regions (see Figure \ref{fig2}). The metric function is thus expressed as:
\begin{eqnarray}\label{eq16}
A_+\left(x\right)=1 + \left(\frac{4m}{\pi{a^3}}\right)\left[xa + \left(x^2+a^2\right)\left(\arctan\left(\frac{x}{a}\right)-\frac{\pi}{2}\right)\right] \qquad x \geq 0, \nonumber \\
A_-\left(x\right)=1 - \left(\frac{4m}{\pi{a^3}}\right)\left[xa + \left(x^2+a^2\right)\left(\arctan\left(\frac{x}{a}\right)+\frac{\pi}{2}\right)\right] \qquad x \leq 0.
\end{eqnarray}

\begin{figure}[htb!]
		\centering
		\mbox{\subfigure[]
			{\label{fig3}
				\includegraphics[scale=0.4]{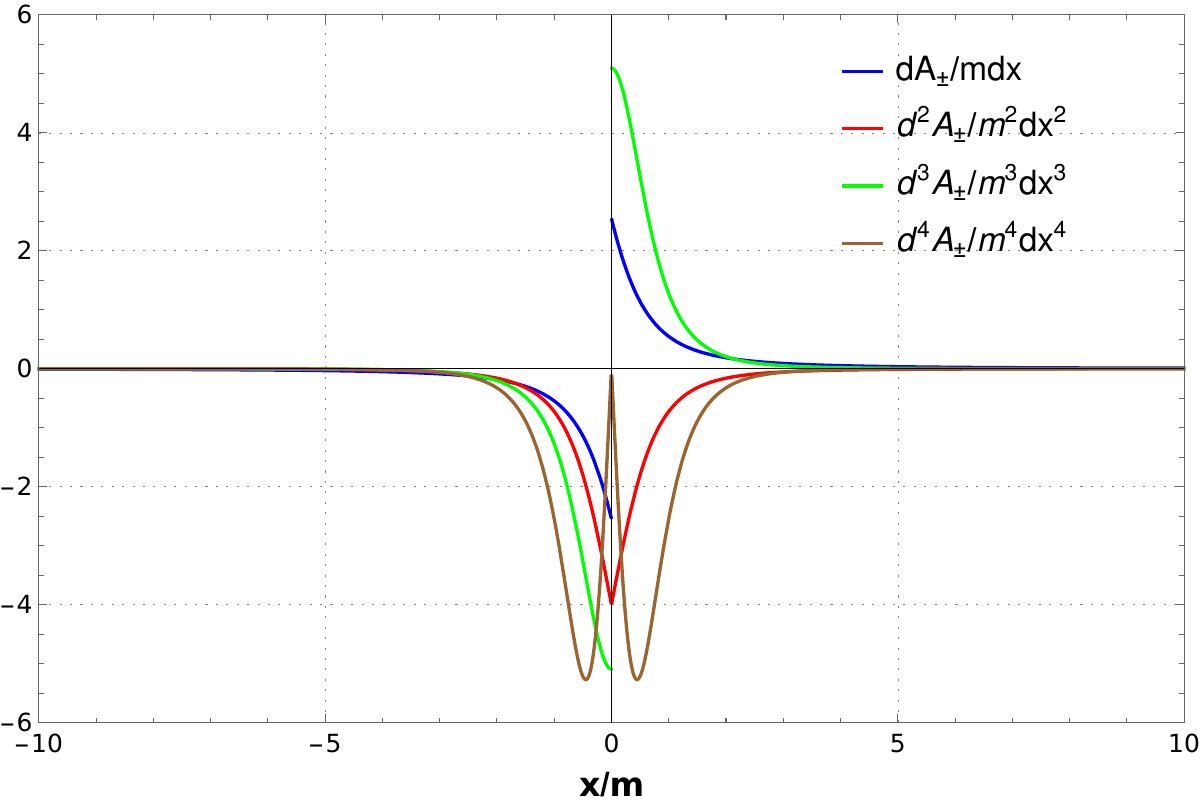}}\qquad
			\subfigure[]
			{\label{fig4}
				\includegraphics[scale=0.4]{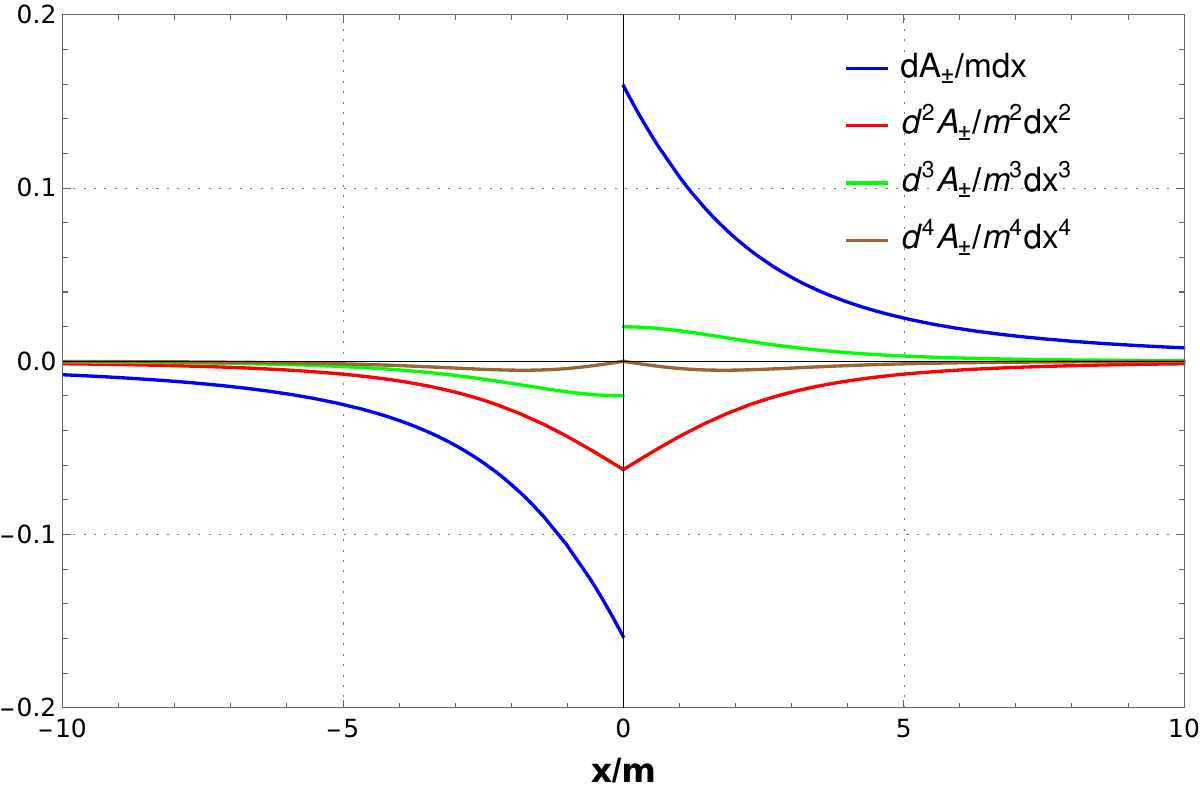}}
}
    \caption{(a) shows the odd and even derivatives of the asymptotically flat function Eq. (\ref{eq15}) for a radius $a=m$ inside the horizon. (b) shows the derivatives for a radius $a=4m$ outside the horizon.}
\label{colagem}
\end{figure}

Figure \ref{colagem} shows curves of derivatives up to fourth order for the metric function Eq. (\ref{eq15}). Figure \ref{fig3} shows the derivatives for a throat radius $a=m$ inside the event horizon. Figure \ref{fig4} displays the derivatives for a radius $a=4m$ outside the event horizon.

The odd derivatives of the metric function Eq. (\ref{eq16}) exhibit discontinuity at the origin, as shown in \ref{colagem}, while even derivatives are continuous, as expected for a smooth function. This arises due to the construction method in Eq. (\ref{eq16}) and implies a spherically symmetric thin shell exists at the junction point $x=0$. Consequently, only traversable wormhole black-bounce solutions are possible, eliminating black hole solutions. This restriction is further examined in Appendix \ref{appendix} and is similar to previous studies \cite{manoel2}.

At this stage, the metric functions have been constructed to meet all necessary conditions. The set of equations of motion, Eqs. (\ref{eq11}-\ref{eq14}), can be rewritten in terms of the metric function $A_\pm(x)$ for each region.

Equation (\ref{eq12}) was used for the area metric function from the original work \cite{matt} to derive the corresponding function $A_\pm(x)$. To obtain the associated scalar field, Eq. (\ref{eq11}) is solved for the $k$-essence field, defined as $X=-\eta A_\pm\phi'^2$ and $F(X)=F_0X^n-2V(\phi)$, where $F_0$ is a constant, $n$ is real, and $V(\phi)$ is the potential. With fixed $n=\frac{1}{3}$ and $\eta=-1$, Eq. (\ref{eq11}) becomes:
\begin{eqnarray}\label{eq21}
{\phi'}_{\pm}= \left(\frac{6}{F_0}\frac{{\Sigma}''}{\Sigma}\right)^\frac{3}{2}A_\pm.
\end{eqnarray}

The above relation is a first order differential equation containing only the metric functions $\Sigma$ and $A_\pm$. Direct integration produces the scalar field $\phi_\pm(x)$, already found in Eq. \eqref{eqphiDsitter}, now, for each region:
\begin{eqnarray}\label{eq22}
\phi_{\pm}\left(x\right)= \frac{D_1}{4a^5}\left[\frac{xa^3}{\Sigma^4}+\frac{3xa}{2\Sigma^2}+\frac{3}{2}\arctan\left(\frac{x}{a}\right)\right] \mp \frac{D_1m}{\pi{a^2}\Sigma^4} - \frac{D_1m}{a^6}\left[\frac{ax}{\Sigma^2}+\arctan\left(\frac{x}{a}\right)\right] \\\nonumber
\pm \left(\frac{2D_1m}{\pi{a^6}}\right)\left[\frac{a^2}{2\Sigma^2}+ \arctan\left(\frac{x}{a}\right)\left(\frac{xa}{\Sigma^2} + \frac{1}{2}\arctan\left(\frac{x}{a}\right)\right)\right],
\end{eqnarray} where $D_1=\left(\frac{6a^2}{F_0}\right)^\frac{3}{2}$ is a constant.

To satisfy the system Eqs. (\ref{eq11}-\ref{eq14}), a scalar potential is required. Eq. (\ref{eq14}) is thus used together with the metric functions $\Sigma$ and $A_\pm(x)$ and the scalar field Eq. (\ref{eq22}) to define the associated potential $V_\pm(x)$:
\begin{eqnarray}\label{eq24}
V_\pm\left(x\right)= A_\pm \frac{{\Sigma}''}{\Sigma} +\frac{1}{\Sigma^2}-\frac{A_\pm'\Sigma'}{\Sigma}-\frac{A_\pm{\Sigma'}^2}{\Sigma^2}.
\end{eqnarray}
The potential in Eq. (\ref{eq24}) is obtained through a procedure analogous to the scalar field definition in Eq. (\ref{eq22}). With some algebraic simplifications, it can be expressed explicitly as:
\begin{eqnarray}\label{eq25}
V_\pm\left(x\right)= \frac{2a^2}{\Sigma^4} \mp \frac{cax}{\Sigma^4}\left(\Sigma^2+2x^2\right) \mp \frac{c}{\Sigma^2}\left(3x^2-a^2\right)\left[\arctan\left(\frac{x}{a}\right)\mp \frac{\pi}{2}\right],
\end{eqnarray}  where $c=\frac{4m}{\pi{a^3}}$ is a combination of constants.

With the scalar potential defined, verification shows all equations of motion are satisfied. In particular, Eq. (\ref{eq13}), which was not used in the derivation, is also satisfied in both regions:
\begin{eqnarray}\label{eq28}
\frac{dV_{\pm}}{dx} + \frac{F_0}{3}\left(\frac{{\phi'_{\pm}}}{\Sigma^2}\right)\left(\sqrt{\frac{F_0\Sigma^5}{6{\Sigma}''}}\right)'=0.
\end{eqnarray}

\begin{figure}[htb!]
		\centering
		\mbox{\subfigure[]
			{\label{escalar1}
				\includegraphics[scale=0.4]{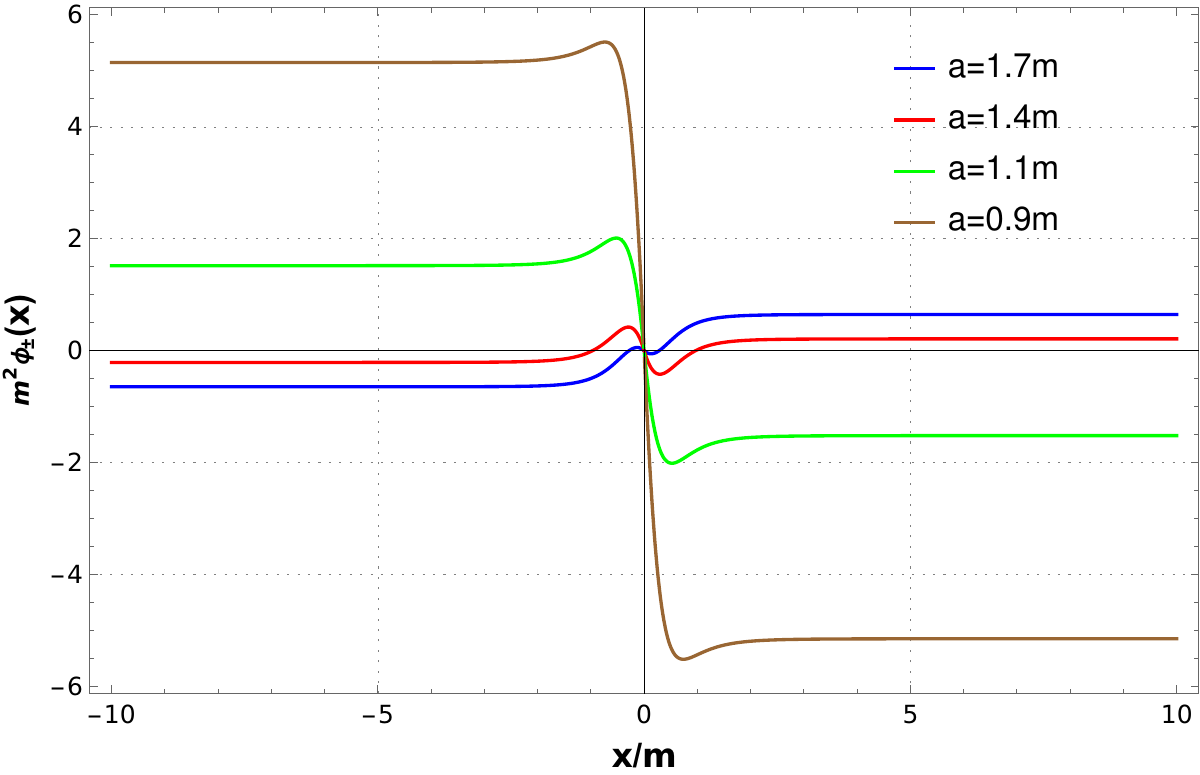}}\qquad
			\subfigure[]
			{\label{escalar2}
				\includegraphics[scale=0.4]{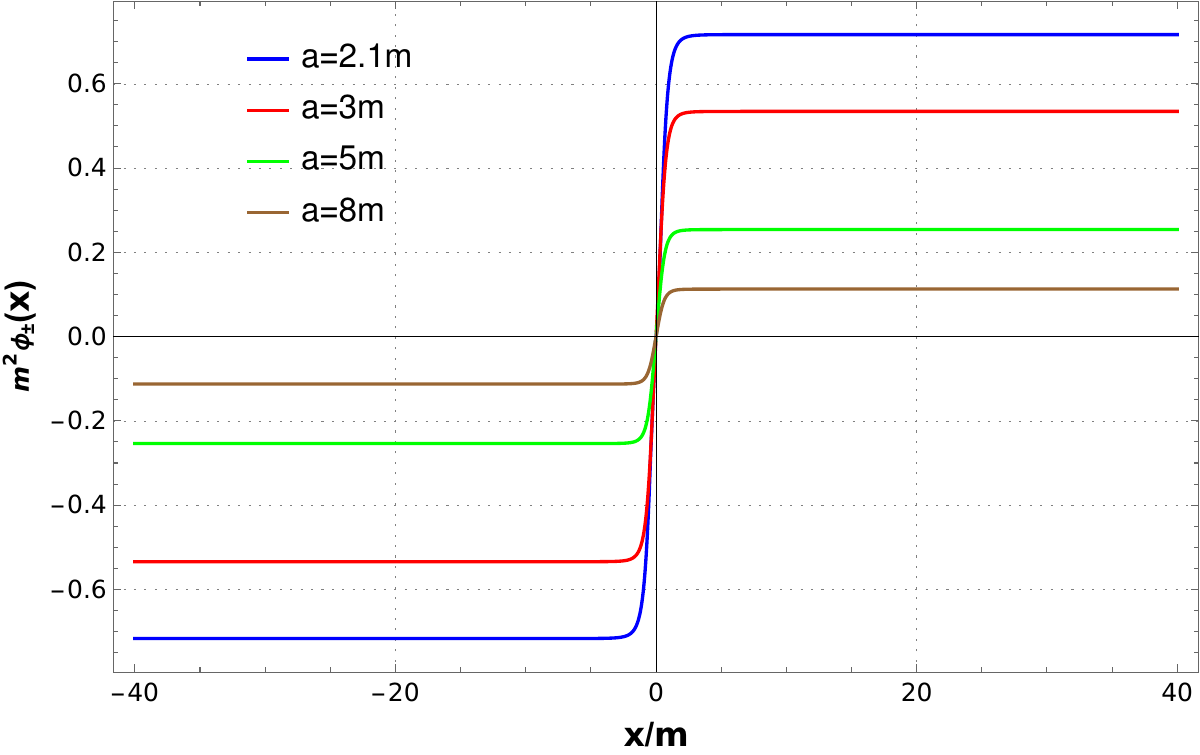}}
}
    \caption{Curves of the scalar field Eq. (\ref{eq22}) for throat radii inside the horizon are displayed in (a), with constant $F_0=1$. In (b), curves are exhibited for radii outside the horizon, also fixing $F_0=1$.
    }
\label{ESCALAR}
\end{figure}

\begin{figure}[htb!]
		\centering
		\mbox{\subfigure[]
			{\label{pote2}
				\includegraphics[scale=0.4]{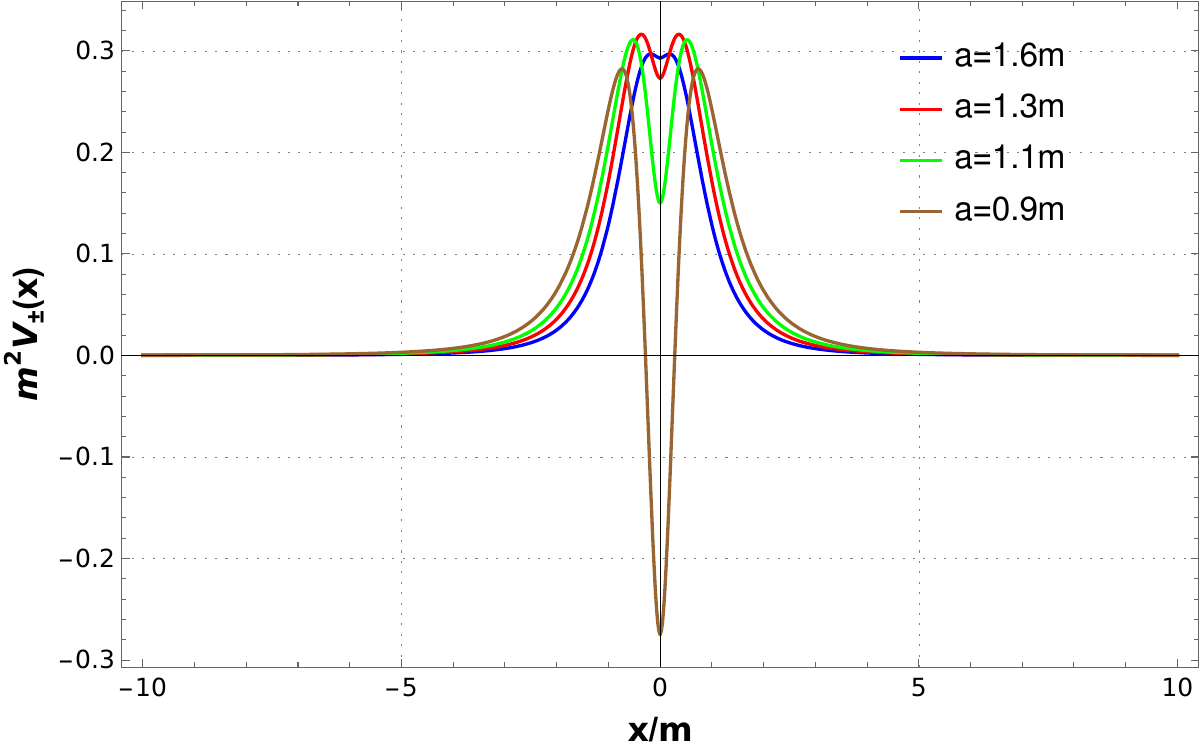}}\qquad
			\subfigure[]
			{\label{pote1}
				\includegraphics[scale=0.4]{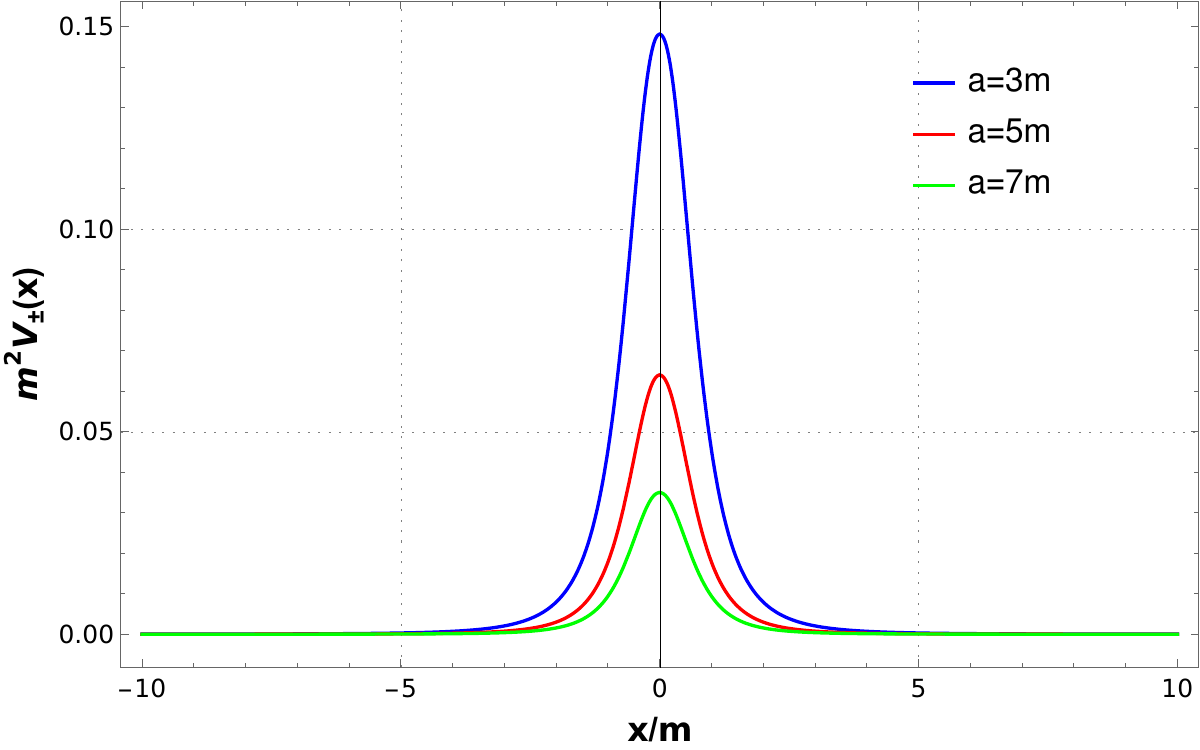}}
}
    \caption{Curves of the potential Eq. (\ref{eq25}) for throat radii inside the horizon are showed in (a). In (b), curves are showed for radii outside the horizon.
    }
\label{POTENCIAL}
\end{figure}

Figures \ref{ESCALAR} and \ref{POTENCIAL} show the scalar field Eq. (\ref{eq22}) and potential Eq. (\ref{eq25}) for various throat radii. The discontinuity and symmetry in the curves reflects the match procedure for the metric function $A_\pm(x)$. Discontinuities in odd derivatives are also showed.

The scalar field exhibits oscillations resulting from interaction with the thin shell at $x=0$ for radii inside the horizon, as shown in \ref{escalar1}. In contrast, the potential acts as a barrier growing near the horizon and decaying at larger radii outside the horizon, as showed in \ref{pote1}. For inside radii in \ref{pote2}, the potential shape takes a form similar to the Pöschl–Teller potential \cite{teller1,teller2,teller3,teller4}.
\section{Geometric quantities }\label{sec4}

With the solutions constructed, the focus now turns to investigating the geometric properties before analyzing the energy conditions. The spherically symmetric line element is defined as:
\begin{eqnarray}\label{eq29}
ds^2= A_\pm\left(x\right)dt^2 -\frac{dx^2}{A_\pm\left(x\right)}  - \Sigma^2\left(x\right)d\Omega^2.
\end{eqnarray}

Constructing the Kretschmann scalar requires the non-zero Riemann tensor components. With the area metric function defined as $\Sigma^2(x) = x^2 + a^2$, the non-vanishing elements are:
\begin{eqnarray}\label{eq30}
\tensor{R}{^{tr}_{tr}}= \frac{A''_\pm}{2}, \quad \tensor{R}{^{\theta\phi}_{\theta\phi}}= \frac{A_\pm\Sigma'^2-1}{\Sigma^2}, \quad
\tensor{R}{^{t\theta}_{t\theta}}=\tensor{R}{^{t\phi}_{t\phi}}= \frac{A'_\pm\Sigma'}{2\Sigma},  \quad  \tensor{R}{^{r\theta}_{r\theta}} =\tensor{R}{^{r\phi}_{r\phi}} = \frac{A'_\pm\Sigma' + 2A_\pm\Sigma''}{2\Sigma}. \nonumber \\
\end{eqnarray}

Using the non-zero Riemann tensor components from Eq. (\ref{eq30}), the Kretschmann scalar $K=\tensor{R}{_{\alpha\beta\mu\nu}}\tensor{R}{^{\alpha\beta\mu\nu}}$ can be constructed in terms of the Riemann tensor as a semi-positive sum of quadratic terms \cite{lobo,Bronnikov2}:
\begin{eqnarray}\label{eq31}
K= 4\left( \tensor{R}{^{tr}_{tr}} \right)^2 + 4\left(\tensor{R}{^{t\theta}_{t\theta}}\right)^2 + 4\left(\tensor{R}{^{t\phi}_{t\phi}}\right)^2+ 4\left(\tensor{R}{^{r\theta}_{r\theta}}\right)^2 + 4\left(\tensor{R}{^{r\phi}_{r\phi}}\right)^2 + 4\left(\tensor{R}{^{\theta\phi}_{\theta\phi}}\right)^2.
\end{eqnarray} 

That imposing the spherical symmetry conditions, can be written in a reduced form, by the expression below:
\begin{eqnarray}\label{eq32}
K= 4\left( \tensor{R}{^{tr}_{tr}} \right)^2 + 8\left(\tensor{R}{^{t\theta}_{t\theta}}\right)^2 + 8\left(\tensor{R}{^{r\theta}_{r\theta}}\right)^2  + 4\left(\tensor{R}{^{\theta\phi}_{\theta\phi}}\right)^2.
\end{eqnarray} 

The Riemann tensor components in Eq. (\ref{eq30}) show the Kretschmann scalar must be defined piecewise due to its dependence on the metric function $A_\pm(x)$. Thus, the Kretschmann scalar is:
\begin{eqnarray}\label{eq33}
K_+\left(x\right)= \frac{\left(\Sigma^2{A''_+}\right)^2 + 2 \left(\Sigma{\Sigma}'{A'_+}\right)^2 + 2\Sigma^2\left({\Sigma}'{A'_+}+ 2A_+{\Sigma}''\right)^2 + 4\left(1-A_+{\Sigma'}^2\right)^2 }{\Sigma^4} \qquad x \geq {0}, \nonumber \\
K_-\left(x\right)= \frac{\left(\Sigma^2{A''_-}\right)^2 + 2 \left(\Sigma{\Sigma}'{A'_-}\right)^2 + 2\Sigma^2\left({\Sigma}'{A'_-}+ 2A_-{\Sigma}''\right)^2 + 4\left(1-A_-{\Sigma'}^2\right)^2 }{\Sigma^4} \qquad x \leq {0}.
\end{eqnarray}
\begin{eqnarray}\label{eq34}
K\left(x\to{0}\right)=  \frac{4\left(3a^2-8am +12m^2\right)}{a^6}
\end{eqnarray}
\begin{figure}[htb!]
\mbox{\subfigure[]
		{\label{KSDESITTER}
			\includegraphics[scale=0.4]{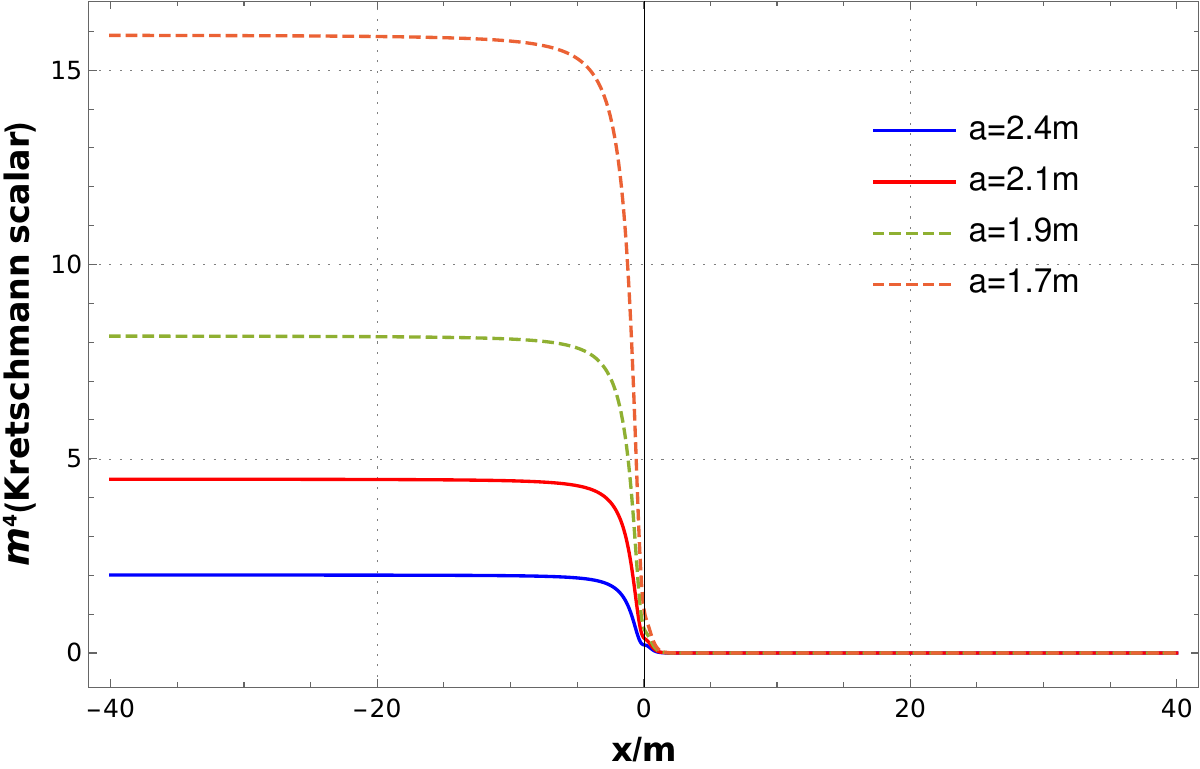}} \qquad
			\subfigure[]
			{\label{K1}
			\includegraphics[scale=0.54]{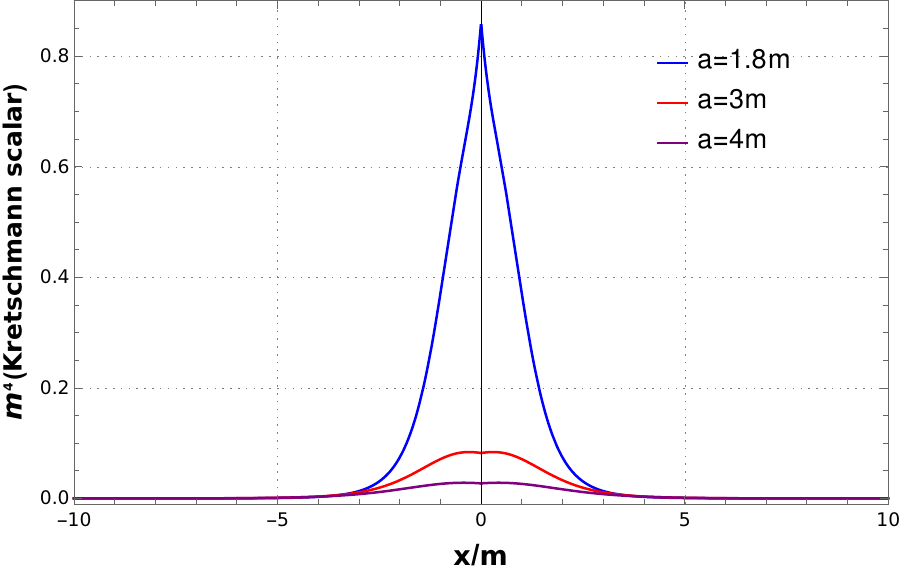}}
}
	\caption{The figure plots on the right represents the Kretschmann scalar for selected throat parameter values $m$, with $a=1.8m$ inside the horizon (blue curve) and $a=3m$, $4m$ outside the horizon (red and purple curves respectively). On the left, we have the Kretschmann scalar for some throat values inside and outside the horizon for the general expression Eq.(\ref{eq15})}
\label{K}
\end{figure}

Note that in equation Eq. (\ref{eq34}), that the Kretschmann scalar is regular in the limit of $x\to{0}$ and therefore, no singularity is present. Likewise, in the limit of $x\to{\pm}\infty$, the scalar goes to zero.

Figure \ref{K1} plots the Kretschmann scalar for throat radii inside and outside the horizon. For $a=1.8m$ within the horizon, Eq. (\ref{eq34}) demonstrates a finite value at the origin. Similarly, the curves for radii outside the horizon also exhibit finite values at the origin.

\subsection{Energy conditions}

Analyzing the null energy conditions requires starting from Einstein's equation \cite{wald}, previously defined in Eq. (\ref{eq1}). This gives the non-zero stress-energy tensor components \cite{inverno} as:
\begin{eqnarray}\label{eq36}
\tensor{T}{^\mu}_{\nu}= {\rm diag}\left[\rho^\phi,-p^\phi_1,-p^\phi_2,-p^\phi_2\right],
\end{eqnarray}
where $\rho^\phi$ is the scalar field energy density, $p^\phi_1$ the radial pressure, and $p^\phi_2$ the tangential pressure. Using the stress-energy tensor diagonal component expressions in Eqs. (\ref{eq4}-\ref{eq5}) for the $k$-essence configuration $n=\frac{1}{3}$ from Eq. (\ref{eq21}) and associated potential Eq. (\ref{eq24}),
\begin{eqnarray}\label{eq37}
\rho^\phi_{\pm}&=& -\frac{F_0}{2}\left[-\eta{A_{\pm}{\left(\phi'_{\pm}\right)}^2}\right]^\frac{1}{3} + V_{\pm}\left(x\right)= -\frac{3{A_{\pm}{\Sigma}''}}{\Sigma} +V_{\pm}\left(x\right), \\\label{eq38} 
p^\phi_{1\pm}&=& - T^{1}_{1} =  \frac{{A_{\pm}{\Sigma}''}}{\Sigma} -V_{\pm}\left(x\right), \\\label{eq39} 
p^\phi_{2\pm}&=&  - T^{2}_{2}= - T^{0}_{0}=- \rho^\phi_{\pm}= \frac{3{A_{\pm}{\Sigma}''}}{\Sigma} - V_{\pm}\left(x\right).
\end{eqnarray}

The defined stress-energy tensor diagonal components are only valid outside the horizon where $A_\pm>0$, with metric signature $(+,-,-,-)$ and $t$ timelike and $x$ spacelike.
 
Inside the horizon, $t$ becomes spacelike and $x$ timelike. The signature changes to $(-,+,-,-)$ with $A_\pm<0$, reversing the coordinate roles. The stress-energy tensor components must then be rewritten as:
\begin{eqnarray}\label{eq40}
\tensor{T}{^\mu}_{\nu}= {\rm diag}\left[-p^\phi_1,\rho^\phi,-p^\phi_2,-p^\phi_2\right],
\end{eqnarray}
and therefore, the equations for energy density, radial pressure, and tangential pressure must be rewritten as:
\begin{eqnarray}\label{eq41}
\rho^\phi_{\pm}&=& -\frac{{A_{\pm}{\Sigma}''}}{\Sigma} +V_{\pm}\left(x\right),\\\label{eq42} 
p^\phi_{1\pm}&=&  \frac{3{A_{\pm}{\Sigma}''}}{\Sigma} - V_{\pm}\left(x\right),\\\label{eq43}
p^\phi_{2\pm}&=& - T^{2}_{2}= - T^{0}_{0}=-\rho^\phi_{\pm}=-\left(-p^\phi_{1\pm}\right)=\frac{3{A_{\pm}{\Sigma}''}}{\Sigma} -V_{\pm}\left(x\right).
\end{eqnarray} 

The constructed geometric quantities depend on the metric function $A_\pm(x)$, so they are defined piecewise. With the defined energy density and pressure components, the energy conditions for black-bounce solutions can now be examined \cite{livro}.

The commonly used energy conditions are inequalities relating the energy density and pressures \cite{livro}:
\begin{eqnarray}\label{eq44}
NEC_{1,2}&=&WEC_{1,2}=SEC_{1,2} \Longleftrightarrow \rho^\phi_{\pm} + p^\phi_{\left(1,2\right)\pm} \geq 0, \\\label{eq45}
SEC_3 &\Longleftrightarrow & \rho^\phi_{\pm} + p^\phi_{1\pm} + 2p^\phi_{2\pm} \geq 0, \\\label{eq46}
DEC_{1,2} &\Longleftrightarrow &  \rho^\phi_{\pm} + p^\phi_{\left(1,2\right)\pm} \geq 0  \qquad    \mbox{and} \qquad \rho^\phi_{\pm} - p^\phi_{\left(1,2\right)\pm} \geq 0 , \\\label{eq47}
DEC_3&=&WEC_{3} \Longleftrightarrow   \rho^\phi_{\pm}  \geq 0 .
\end{eqnarray}

The energy conditions can be explicitly expressed in terms of the metric functions by substituting the stress-energy tensor components from Eqs. (\ref{eq37}-\ref{eq39}) into the defining inequalities Eqs. (\ref{eq44}-\ref{eq47}).

This gives the energy conditions in the timelike region outside the event horizon where $A_\pm > 0$ as:
\begin{eqnarray}\label{eq48}
NEC^\phi_{1}&=&WEC^\phi_{1}=SEC^\phi_{1} \Longleftrightarrow  -\frac{2A_{\pm}\Sigma''}{\Sigma} \geq 0, \\\label{eq49}
NEC^\phi_{2}&=&WEC^\phi_{2}=SEC^\phi_{2} \Longleftrightarrow  0, \\\label{eq50}
SEC^\phi_3 & \Longleftrightarrow &  \frac{4{\Sigma}''{A_{\pm}}}{\Sigma} -2V_{\pm}\left(x\right)  \geq 0, \\\label{eq51}
DEC^\phi_{1} & \Longleftrightarrow & -\frac{4{\Sigma}''{A_{\pm}}}{\Sigma}                                + 2V_{\pm}\left(x\right) \geq 0, \\\label{eq52}
DEC^\phi_{2} & \Longleftrightarrow & -\frac{6{\Sigma}''{A_{\pm}}}{\Sigma}                                + 2V_{\pm}\left(x\right) \geq 0, \\\label{eq53}
DEC^\phi_{3}&=&WEC^\phi_{3}  \Longleftrightarrow   -\frac{3{A_{\pm}{\Sigma}''}}{\Sigma} +V_{\pm}\left(x\right) \geq 0.
\end{eqnarray}
Likewise, the energy conditions inside the horizon where $t$ is spacelike are obtained by substituting the stress-energy tensor components from Eqs. (\ref{eq41}-\ref{eq43}) into the inequalities Eqs. (\ref{eq44}-\ref{eq47}). This gives the energy conditions for $A_\pm < 0$ as:
\begin{eqnarray}\label{eq54}
NEC^\phi_{1}&=&WEC^\phi_{1}=SEC^\phi_{1} \Longleftrightarrow  \frac{2A_{\pm}\Sigma''}{\Sigma} \geq 0, \\\label{eq55}
NEC^\phi_{2}&=&WEC^\phi_{2}=SEC^\phi_{2} \Longleftrightarrow   \frac{2A_{\pm}\Sigma''}{\Sigma} \geq 0, \\\label{eq56}
SEC^\phi_3 & \Longleftrightarrow &  \frac{8A_{\pm}\Sigma''}{\Sigma} -2V_{\pm}\left(x\right)\geq 0, \\\label{eq57}
DEC^\phi_{1} & \Longleftrightarrow & -\frac{4A_{\pm}\Sigma''}{\Sigma} +2V_{\pm}\left(x\right) \geq 0, \\\label{eq58}
DEC^\phi_{2} & \Longleftrightarrow & -\frac{4A_{\pm}\Sigma''}{\Sigma} +2V_{\pm}\left(x\right) \geq 0, \\\label{eq59}
DEC^\phi_{3}&=&WEC^\phi_{3} \Longleftrightarrow   -\frac{A_{\pm}\Sigma''}{\Sigma} +V_{\pm}\left(x\right) \geq 0.
\end{eqnarray}

{Equations (\ref{eq48}-\ref{eq54}) demonstrate that the null energy condition $(NEC^{\phi}_{1})$ is violated both inside and outside the event horizon. Likewise, $NEC^{\phi}_{2}$ given by Eq.(\ref{eq49}) is satisfied outside the horizon but violated inside according to Eq.(\ref{eq55}). Since $DEC^{\phi}_{2}$ is connected to $NEC^{\phi}_{2}$, it is also violated within the horizon through Eq. (\ref{eq58}). Similarly, $DEC^{\phi}_{1}$ is violated both outside and inside the horizon, which is tied to the violation of $(NEC^{\phi}_{1})$.}

\begin{figure}[htb!]
		\mbox{\subfigure[]
			{\label{DEC2a}
				\includegraphics[scale=0.54]{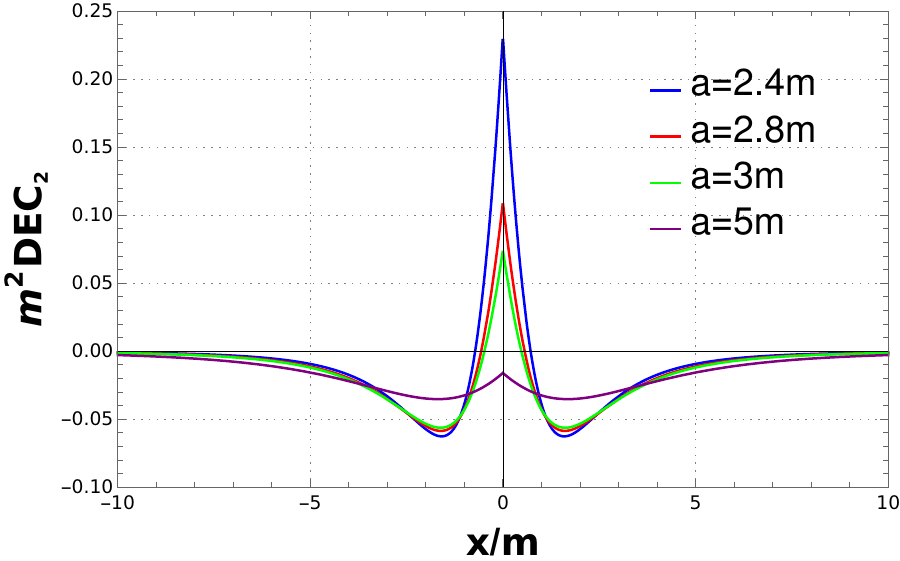}}\qquad
			\subfigure[]
			{\label{DEC3}
				\includegraphics[scale=0.54]{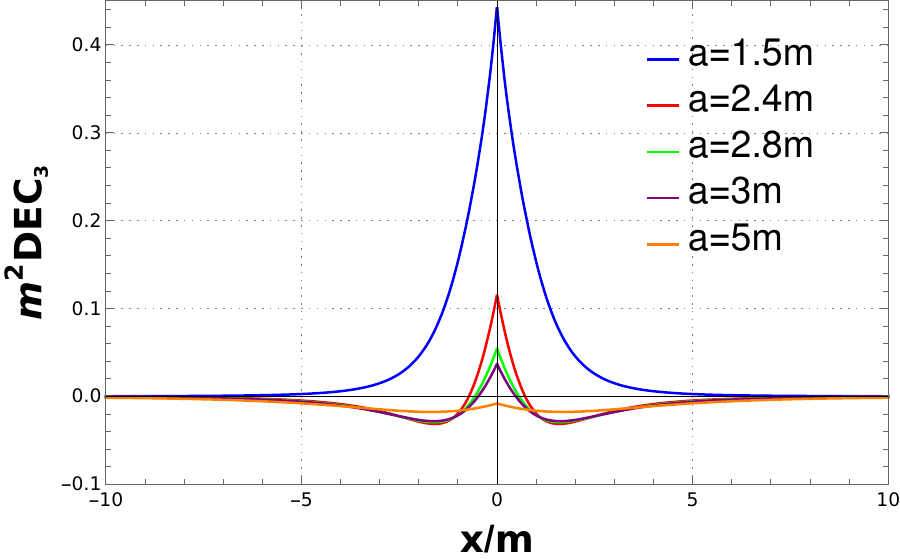}} 
}
    \caption{DEC plot relating energy density and tangential pressure, for radii inside and outside the event horizon.
    }
\label{DEC2}
\end{figure}

\begin{figure}[htb!]
				{\includegraphics[scale=0.7]{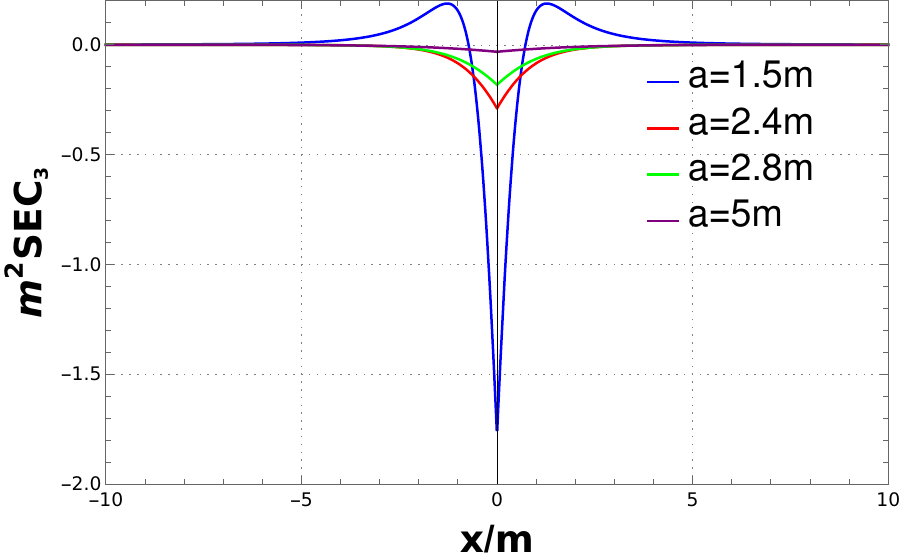}}
    \caption{Strong energy condition $\left(SEC\right)$ plot combining energy density and all pressure components, for various radii inside and outside the horizon.
    }
\label{SEC3}
\end{figure}

Complementarily, Fig. \ref{DEC2a} exhibits $DEC^\phi_2$ violation for all radii outside the horizon. However, $DEC^\phi_3$ violates outside but satisfies inside the horizon (Fig. \ref{DEC3}). Finally, $SEC^\phi_3$ violates inside and outside, as shown in Fig. \ref{SEC3}.

\section{Conclusion}\label{sec5}

The present analysis utilizes the $k$-essence field equations describing a phantom scalar field to construct black-bounce solutions not possible with an ordinary scalar field. It should be noted that k-essence does not constitute a modified theory of gravity; rather, it introduces a scalar field through a non-standard kinetic term. The analysis begins with the areal metric function $\Sigma^2 = x^2 + a^2$ containing a throat radius $a$ as in the original black-bounce proposals \cite{matt}. The corresponding metric function in $k$-essence theory is derived by applying boundary conditions to obtain an asymptotically flat spacetime. This defines the full metric and enables study of the black-bounce structures.

The analysis attempts to satisfy the equations of motion with only a kinetic term for the scalar field. However, this is insufficient, requiring introduction of a scalar potential as well. Analytical expressions for the scalar field and necessary potential are derived, with the full set of equations satisfied. The possibility of using alternative black-bounce throat metric functions, as studied in \cite{manoel1}, is also examined but leads to algebraically intractable solutions.

With the derived analytical metric function and known black-bounce throat function, the Kretschmann scalar is verified to be regular at the origin for radii inside and outside the horizon. The mixed stress-energy tensor components are defined on each side of the horizon, with the roles of $t$ and $x$ reversed. Analysis of the energy conditions shows violation of the null energy condition ($NEC^{\phi}_{1}$) inside and outside the horizon, consistent with other black-bounce solutions. Violation of the null energy condition is the main ingredient for building regular black-bounce geometries.

{As is well known, in general relativity the strong energy condition $(SEC_3)$ is typically violated within the event horizon for regular black hole solutions, while the weak energy condition $(WEC)$ can be violated throughout spacetime in some cases \cite{matt,manoel2}. However, the solution presented in this work exhibits different behavior, with the strong energy condition $(SEC^{\phi}_{3})$ being violated both outside and inside the event horizon, as shown in Fig. \ref{SEC3}. Meanwhile, the weak energy condition $(WEC^{\phi}_{3})$ is violated outside the horizon but satisfied inside, as depicted in Fig. \ref{DEC3}.}

{An interesting observation is that due to the visual form that the potential takes in the figures \ref{POTENCIAL} we may be indicating a possible stability of the solutions when subjected to radial perturbations. Considering the possibility of having normal and quasi-normal modes \cite{teller1,teller2,teller3,teller4}}.
	
\begin{acknowledgments}
We thank CNPq, CAPES and FAPES for financial support. The authors thank M. E. Rodrigues; M. V. S. Silva and E. L. Martins for their fruitful discussions.
\end{acknowledgments}
	
%\section*{Data availability}

\appendix
\section{Thin shell in back-bounce solution}\label{appendix}

This section is reserved to demonstrate the sign of the mass of the surface at the point where the metric function was matched (Eq. \ref{eq16}). For the line element contained in Eq. (\ref{eq29}), the following coordinate transformation will be considered, $r^2 = x^2 + a^2$ which transforms into an equivalent line element and was adopted in \cite{thin1,thin2}. It is emphasized that in this last section, the same signature of the metric will be used as in the cited works above so that the results can be better compared. In this way, the line element Eq. (\ref{eq29}) gets rewritten as
\begin{eqnarray}\label{apen1}
ds^2=  A_{\pm}\left(r\right)dt^2 - \frac{dr^2}{A_{\pm}\left(r\right)\left(1-\frac{a^2}{r^2}\right)} - r^2d\Omega^2.
\end{eqnarray} 
The metric function $A_{\pm}\left(r\right)$ is defined in terms of the new coordinate as:
\begin{eqnarray}\label{apen2}
A_{\pm}\left(r\right)= 1\pm\frac{4m}{\pi{a^3}}\left[a\left(\sqrt{r^2-a^2}\right)+r^2\left(\arctan\left(\frac{\sqrt{r^2-a^2}}{a}\right)\mp{\frac{\pi}{2}}\right)\right].
\end{eqnarray}

In the original metric (Eq. \ref{eq29}), the coordinates $x$ and $t$ are defined over the entire space $x\in$ $\left(-\infty,+\infty\right)$ and $t\in$ $\left(-\infty,+\infty\right)$ \cite{matt}. In the new coordinates (Eq. \ref{apen1}), the temporal part $t$ retains the same range, but the radial part domain is modified to $r\in$ $\left(a, +\infty\right)$.
The line element describing the thin shell is given by
\begin{eqnarray}\label{apen3}
ds^2=  d\tau^2 - R^2\left(\tau\right)d\Omega^2,
\end{eqnarray} 
where the parameter $\tau$ corresponds to the proper time for an observer in the shell.

To compute the extrinsic curvature, the 4-velocity vector $U^{\mu} = \left(\frac{dt}{d\tau},\frac{dR(\tau)}{d\tau},0,0\right)$ and normal vector $n^{\mu}$ to the hypersurface are first defined. The 4-velocity vector can be expressed in terms of the metric components in Eq. (\ref{apen1}) as
\begin{eqnarray}\label{apen4}
U^{\mu}_\pm=\pm \left[\sqrt{\frac{\left(1+g_{11}\dot{R}^2\right)}{g_{00}}},\dot{R},0,0\right],
\end{eqnarray} on what $\dot{R}=\frac{dR}{d\tau}$ and $U\mu{U^{\mu}}=1$. 

In the same way, we will define the normal vector to the surface. For this, we will need to perform a parameterization in terms of the intrinsic coordinates $\xi^{i}=\left(\tau,\theta,\phi\right)$  Eq. (\ref{apen3}).
Therefore, the parameterization is defined as $f(x^{\mu}(\xi^i)) = r - R(\tau) = 0$, and the normal unit 4-vector is given by the expression
\begin{eqnarray}\label{apen5}
n_\mu= \frac{\nabla_\mu{f}}{||\nabla{f}||}= \pm \left| g^{\alpha\beta}\frac{\partial{f}}{\partial{x^{\alpha}}}\frac{\partial{f}}{\partial{x^{\beta}}}\right|^{-\frac{1}{2}} \frac{\partial{f}}{\partial{x^{\mu}}}.
\end{eqnarray} The normal vector is unitary $n_\mu{n^\mu}=-1$ and orthogonal to the vectors tangent to the surface $n_{\mu}e^{\mu}_{i}=n_{\mu}\left( \frac{\partial{x^{\mu}}}{\partial{\xi^{i}}}\right)=0$. Therefore, the normal vector written in terms of the components of the metric Eq. (\ref{apen1}) is given by
\begin{eqnarray}\label{apen6}
n^{\mu}_\pm= \pm \left[\dot{R}\sqrt{-\frac{g_{11}}{g_{00}}},\sqrt{-g^{11}+\dot{R}^2},0,0\right].
\end{eqnarray}
With the constructed normal vector $n^{\mu}$ and 4-velocity vector $U^{\mu}$, the extrinsic curvature can be defined as
\begin{eqnarray}\label{apen7}
K^{\pm}_{ij}=-n_{\mu}\left[\frac{\partial^2{x^\mu}}{\partial\xi^{i}\partial\xi^{j}}+ \Gamma^{\mu\pm}_{\alpha\beta}\frac{\partial{x^{\alpha}}}{\partial{\xi^{i}}}\frac{\partial{x^{\beta}}}{\partial{\xi^{j}}}\right].
\end{eqnarray}
The $\theta\theta$ component of the extrinsic curvature is computed, as it is related to the surface energy density.
Thus, its explicit form is given by
\begin{eqnarray}\label{apen8}
K^{\theta\pm}_{\theta}= \pm \frac{1}{R}\left[A_{\pm}\left(1-\frac{a^2}{R^2}\right) + \dot{R}^2\right]^{\frac{1}{2}}.
\end{eqnarray} 

\subsection{Lanczos Equation}

The discontinuity across the thin shell is characterized by the difference in extrinsic curvature outside and inside, $k_{ij} = K^+{ij} - K^-{ij}$. The Einstein equation in the interior spacetime yields the Lanczos equation:
\begin{eqnarray}\label{apen9}
S^{i}_{j}= -\frac{1}{8\pi}\left(k^{i}_{j}-\delta^{i}_{j}k^{k}_{k}\right),
\end{eqnarray}
where $S^i_j$ are the non-zero components of the surface stress-energy tensor, $S^i_j = \text{diag}(-\sigma, \mathcal{P}, \mathcal{P})$. Here, $\sigma$ is the surface energy density and $\mathcal{P}$ is the pressure. The $\tau\tau$ component of the Lanczos equation yields the surface energy density:
\begin{eqnarray}\label{apen10}
\sigma= -\frac{1}{4\pi}k^{\theta}_{\theta}=-\frac{1}{2\pi{R}}\left[A_{\pm}\left(1-\frac{a^2}{R^2}\right) + \dot{R}^2\right]^{\frac{1}{2}}.
\end{eqnarray}

At the junction point $x=0$, the metric function takes the Simpson-Visser form $A_{\pm} = \left(1 - \frac{2m}{R}\right)$ \cite{matt}. For a static shell with $\dot{R} = 0$, the energy density in Eq. (\ref{apen10}) becomes
\begin{eqnarray}\label{apen11}
\sigma =-\frac{1}{2\pi{R}}\left[\left(1-\frac{2m}{R}\right)\left(1-\frac{a^2}{R^2}\right)\right]^{\frac{1}{2}}.
\end{eqnarray}

{Therefore, by analyzing the expression for the static energy density in Eq. (\ref{apen11}), we see that the product of the terms inside the square root imposes two constraints to be positive: $R > a$ and whether the shell is inside or outside the event horizon. Note that for any shell value greater than the throat radius $a$, the second term inside the square root is always positive. However, the first term inside the square root depends on whether the shell is inside $A_{\pm}<0$ or outside $A_{\pm}>0$ the event horizon. This implies that only traversable wormhole solutions are allowed.}
 
{The static energy density in Eq. (\ref{apen11}) can be analyzed generally, without requiring evaluation specifically at the junction surface. Notably, the metric function in Eq. (\ref{apen2}) is positive for throat radii outside the horizon, $a > 2m$, and negative inside, $a < 2m$ (Fig. \ref{fig2}). With the surface mass defined as $m_s = 4\pi R^2\sigma$ and the negative energy density $\sigma$, the surface mass is also negative. This signifies violation of the energy conditions at the junction surface.}

%\clearpage

%\newpage

%Data sharing not applicable to this article as no datasets were generated or analysed during the current study.

\clearpage

\nocite{*}
%\bibliography{electrovac}
		
\end{document}